\documentclass[preprint]{aastex}
\usepackage{spr-astr-addons}
\usepackage{url}
\usepackage{rotating}

\RequirePackage{color}

\begin{document}

\title{Application of the MST clustering to the high energy $\gamma$-ray sky.\\
II - Possible detection of $\gamma$-ray emission from blazar candidates in the 1WHSP sample} 

%% Running heads
\shorttitle{Minimum Spanning Tree cluster analysis of 1WHSP sources.}
\shortauthors{R. Campana et al.}

%% Author and Affiliations
\author{R. Campana}
\affil{INAF--IASF-Bologna, via Piero Gobetti 101, I-40129, Bologna, Italy}
\and
\author{E. Massaro}
\affil{INAF--IAPS, via del Fosso del Cavaliere 100, I-00133, Roma, Italy \\ and \\ In Unam Sapientiam, Roma, Italy}
\and
\author{E. Bernieri}
\affil{INFN--Sezione di Roma Tre, via della Vasca Navale 84, I-00146 Roma, Italy}

\begin{abstract}
We present the results of a photon cluster search in the 7-years \emph{Fermi}-Large Area Telescope 
extragalactic Pass 8 $\gamma$-ray sky by means of the Minimum Spanning Tree (MST) algorithm, 
at energies higher than 10 GeV.
We found 16 clusters of photons, corresponding to candidate $\gamma$-ray sources, located very close to 
infrared-selected sources in the 1WHSP (WISE High Synchrotron Peaked) sample, and therein
classified as either ``new'' or ``candidate'' blazars.
In this paper some properties of the MST clusters and of the associated sources are presented.
\end{abstract}

  
\section{Introduction}\label{s:introduction}

The \emph{Fermi}-Large Area Telescope (LAT) sky survey at $\gamma$-ray energies has shown 
that blazars constitute the largest class of extragalactic high energy sources 
\citep[see, for instance, the 2LAC and 3LAC catalogues,][]{ackermann11, ackermann15}.
This finding has triggered some observational campaigns for discovering possible counterparts
to unassociated $\gamma$-ray sources, aiming to extend the known blazar population.
As discussed for example in the recent review by \cite{massaro16}, 
these researches on the extragalactic $\gamma$-ray sky are important to evaluate the 
contribution of these sources to the high energy background and to establish some constraints 
to the dark matter \citep{ajello15}.
Moreover, some of the newly discovered faint blazars are expected to be located at high 
redshift. The study of their spectra in the GeV band would provide useful information on 
the extragalactic background light \citep[EBL, see, for instance,][]{ackermann16}.

Searches aimed to the discovery of BL Lac objects are particularly interesting because this
type of blazars is generally characterized by hard spectra \citep{ackermann16} and therefore they 
are the majority of the extragalactic sources detected at the highest energies. 
In the last years searches for BL Lacs based on multifrequency approaches have provided new 
rich samples of confirmed and candidate sources like the WISE Blazar Radio-Loud Sources 
\citep[WIBRaLS,][]{dabrusco14} and the subsequent WISE High Synchrotron Peaked 
\citep[1WHSP,][]{arsioli15} catalogues. Both are based on a colour selection from the infrared 
photometric database of the WISE mission \citep{wright10}
according the criteria for the BL Lacs established by \cite{massaro11,massaro13}.
The 1WHSP catalogue includes 992 sources which were selected to be confirmed and candidate
HSP (High Synchrotron Peaked) blazars, i.e. with or without available optical spectra allowing
a safe classification, and therefore one can expect that they should be detected 
up to the highest $\gamma$-ray energies: 299 of them,in fact, are known confirmed emitters of GeV 
photons, based on Fermi-LAT catalogues, and 36 have already been detected in the TeV band.

In a preliminary work \citep[][hereafter Paper I]{campana15} the Minimum 
Spanning Tree (MST) algorithm was successfully applied to the search of 
new spatial clusters of $\gamma$-rays 
which could be an indication for localized high energy sources, and illustrated how 
MST is useful for finding clusters having a small number of photons, but likely 
related to pointlike sources.

Paper I was based on a \emph{Fermi}-LAT Pass 7 dataset spanning 6.3 years. 
However, the new response functions introduced with the Pass 8 reprocessing \citep{atwood13}
ensure an improved event reconstruction, leading to better background rejection, 
effective area and angular resolution.

Therefore, in this paper we report the results of a new MST analysis for detecting photon 
clusters in the 7-years \emph{Fermi}-LAT Pass 8 sky at energies higher than 10 GeV for Galactic 
latitudes higher than 25$^{\circ}$, where we found 16 clusters located very close 
to 1WHSP sources, classified in this catalogue as either \emph{New}  (i.e. not already reported
as blazars) or \emph{Candidate} blazars. 
In other words, the identification of these infrared-selected sources is still not definitive, 
and the $\gamma$-ray results shown in this work could help to clarify their identity as blazars.

In a companion paper \citep{campana16} we will instead focus on the analysis of clusters
matching sources already classified as blazars and reported in the Roma BZCAT catalogue, extending the results of Paper I.

\section{Photon cluster detection by means of the MST algorithm}

The topometric Minimum Spanning Tree algorithm searches for clusters in a 
field of points. 
A description of the method is presented in Paper I, and a more detailed 
discussion of its statistical properties can be found in \cite{campana08,campana13}. 
Therefore we summarize here only its principal characteristics, for the sake of completeness. 

Given a two-dimensional set of $N$ points (\emph{nodes}) one can define a set
$\{\lambda_i\}$ of weighted edges connecting them. 
The MST is the unique graph without closed loops (a \emph{tree}) that connects all the 
nodes with the minimum total weight, $\min [\Sigma_i \lambda_i]$. 
In our case, where the dimensions are the photon arrival coordinates on a celestial sphere,
the edge weights are the angular distances. 

After the computation of the MST, a set of subtrees
corresponding to photon clusters is extracted by means of a 2-step \emph{primary selection}.
The first step (\emph{separation}) removes all the edges having a length $\lambda > \Lambda_\mathrm{cut}$, 
the separation value, usually defined in units of the mean edge length 
$\Lambda_m = (\Sigma_i \lambda_i)/N$. This results in a set of disconnected sub-trees.
The second step (\emph{elimination}) removes all the sub-trees having a number of nodes 
$N \leq N_{cut}$, 
leaving only the clusters having a size over a properly fixed threshold. 
The remaining set of sub-trees provides a first list of clusters. 

A \emph{secondary} selection is then applied to extract the most robust candidates 
for $\gamma$-ray sources.
In Campana et al. (2013) a suitable parameter for this selection was introduced, the 
\emph{magnitude} of the cluster, defined as
$M_k = n_k g_k  $,
where $n_k$ is the number of nodes in the cluster $k$ and the \emph{clustering parameter} $g_k$ 
is the ratio between $\Lambda_m$ and $\lambda_{m,k}$, the mean length of the $k$-th cluster edges. 
Comparative tests performed in simulated and real \textit{Fermi}-LAT fields \citep[see][]{campana13} 
showed that $\sqrt{M}$ has a linear correlation with other statistical 
significance parameters and that it can be a good estimator of the significance of a MST 
cluster.
Usually \citep[][Paper I]{campana13} a lower threshold of $M \sim 20$ is used.

To have an indication if a cluster is compatible with a $\gamma$-ray source two quantities can be computed,
namely the centroid coordinates, obtained by means of a weighted mean of the photons' coordinates 
\citep[see][]{campana13} and the radius of the circle centred at the centroid and containing 
the 50\% of photons in the cluster, the \emph{median radius} $R_m$.
For a cluster likely associated with a genuine pointlike 
$\gamma$-ray source, the latter should be smaller than or comparable to the 68\% containment radius of 
instrumental Point Spread Function (PSF).
This radius varies from 0\fdg25 at 3 GeV to 0\fdg12 at 10 GeV in the case of 
front-converting events
\citep{ackermann13b}, using the latest instrumental response files\footnote{\url{http://www.slac.stanford.edu/exp/glast/groups/canda/lat_Performance.htm}}  \citep[Pass 8,][]{atwood13}. 
We also expect that the angular distance beetween the positions of the cluster centroid 
and the possible optical counterpart are lower than the latter value.

\begin{table*} 
\caption{Coordinates and main properties of MST clusters detected at energies
higher than 10 GeV ($\Lambda_\mathrm{cut} =$ 0.7). 
The second and third columns give the J2000 coordinates of MST cluster centroids; angular distances $\Delta \theta$ from the catalog position of 1WHSP associations are computed from these values.
Blazar types N and C are for \emph{New} and \emph{Candidate}, respectively (see text for details).}
\label{t:sourcelist}
\centering
\begin{tabular}{crrcrrcccl}
\hline
MST	cluster	& RA	& DEC	&	1WHSP	source& $\Delta \theta$ & $n$ & $g$~~~ & $M$~~~ & $R_m$ &   Type \\
	& J2000	& J2000	& 	& $'$ 			  &     &        &        &  deg  &         \\

\hline
MST~0009$-$4316 &     2.412 &  $-$43.317 &  J000949.74$-$431650 & 2.95 &   7 &  3.63 &   25.41 &    0.054 & N \\   
MST~0029$+$2053 &     7.399 &   20.874   &  J002928.60+205333   & 2.01 &   7 &  5.25 &   36.75 &    0.047 & C \\   
MST~0102$-$2002 &    15.675 &  $-$20.021 &  J010250.93$-$200158 & 2.22 &   8 &  4.87 &   38.96 &    0.071 & N \\ 
MST~0241$-$3041 &    40.263 &  $-$30.640 &  J024115.48$-$304140 & 4.21 &   9 &  2.91 &   26.19 &    0.101 & N \\ 
MST~0328$-$5716 &    52.261 &  $-$57.217 &  J032852.68$-$571605 & 3.35 &   6 &  3.75 &   22.50 &    0.080 & N \\ 
MST~0507$-$3346 &    76.864 &  $-$33.782 &  J050727.27$-$334635 & 0.33 &  12 &  3.71 &   44.52 &    0.075 & C \\ 
MST~0812$+$2821 &   123.155 &   28.363   &  J081231.25+282056   & 1.55 &   7 &  2.94 &   20.58 &    0.042 & N \\ 
MST~1033$+$3708 &   158.419 &   37.139   &  J103346.39+370824   & 1.17 &   7 &  3.04 &   21.28 &    0.059 & N \\
MST~1043$+$0054 &   160.690 &    0.929   &  J104303.84+005420   & 4.77 &  10 &  3.07 &   30.67 &    0.089 & C \\ 
MST~1432$+$7644 &   217.767 &   76.730   &  J143211.62+764355   & 3.88 &   8 &  2.60 &   20.80 &    0.111 & C \\ 
MST~1550$-$0822 &   237.718 &   $-$8.360 &  J155053.27$-$082246 & 1.20 &   6 &  6.27 &   37.62 &    0.022 & C \\ 
MST~1623$+$0857 &   245.866 &    8.959   &  J162330.55+085724   & 0.69 &   8 &  3.68 &   29.44 &    0.063 & N \\ 
MST~1626$+$6300 &   246.633 &   62.983   &  J162646.04+630048   & 2.43 &   9 &  4.52 &   40.68 &    0.067 & C \\ 
MST~1640$+$6852 &   250.099 &   68.835   &  J164014.94+685234   & 2.59 &   7 &  4.10 &   28.70 &    0.047 & C \\ 
MST~2002$-$5736 &   300.543 &  $-$57.639 &  J200204.18$-$573645 & 1.79 &   8 &  4.47 &   35.76 &    0.037 & C \\ 
MST~2040$-$4620 &   309.997 &  $-$46.430 &  J204006.61$-$462017 & 5.64$^{(1)}$ &   8 &  2.87 &   22.96 &    0.131 & C \\  
\hline
\end{tabular}
\flushleft{($^{1}$) This angular separation decreases to 2\farcm35 for the cluster found at energies $>$7 GeV.}\\
\end{table*}

\section{The 1WHSP subsample}

The 1WHSP sample \citep{arsioli15} was selected applying a method that 
combines AllWISE survey data \citep{cutri13} with multi-frequency selection 
criteria to extract confirmed and candidate HSP (High Synchrotron Peaked) sources.
We selected a subsample from the entire 1WHSP list (992 sources) by excluding  
418 sources already reported in the 5th Edition of the Roma-BZCAT catalogue 
\citep[hereafter 5BZCAT,][]{massaro14,massaro15}
and those for which a possible counterpart in the \emph{Fermi}-LAT catalogues 
(1FGL, \citealt{abdo10}, 2FGL, \citealt{nolan12}, and 3FGL, \citealt{acero15}) 
is indicated (299 sources).
Finally, we excluded all the sources with an absolute Galactic latitude lower than 25\degr.
The resulting 1WHSP subsample contains 440 sources: 309 are indicated as \emph{Candidate}
blazars, 124 as \emph{New} blazars, 6 as \emph{Known} blazars and 1 as \emph{Sedentary} blazar \citep[i.e.
belonging to the Sedentary Survey;][]{giommi99,giommi05}.
Sources reported as \emph{New} blazars are objects with available spectra in the literature or 
in large databases (like SDSS or 6dF), while for the \emph{Candidates} no safe optical
spectroscopic information was found.
Note that also in the 1WHSP list, similarly to the 5BZCAT, there is a north-south asymmetry in
the number (and density) of sources due to the available surveys' data.
Sources with a positive declination are 529 against 463 with negative values,
but those with declination lower than $-$40\degr, the South NVSS limit, are only 125 while in
the corresponding North region are 209.
This asymmetry, however, is not present in our subsample that has 237 sources
(54\%) with negative declinations.

\section{The MST cluster populations}

We searched for clusters of $\gamma$-ray photons at energies higher than 10 GeV 
detected by \emph{Fermi}-LAT (using 7 years of archival Pass 8 data, spanning from 2008 
August 4 to 2015 August 4) by means of MST. 
The Galactic belt up to a latitude $|b| \leq 25\degr$ was excluded to reduce the
possibility of finding clusters originated by local high background fluctuations.
Each of these two broad spherical regions was then divided into ten smaller subregions
and MST was applied in each of them.
The parameters of the primary selection of clusters were $N_\mathrm{cut} = 4$
and $\Lambda_\mathrm{cut} = 0.7 \Lambda_{m}$; then a rather severe secondary selection was
applied with $M > 20$.
A sample of 921 clusters was obtained, of which 716 have a firm 3FGL counterpart, and
165 a 1FHL counterpart (five of which are not in the 3FGL catalogue).
From the residual clusters we sorted out 10 more corresponding to sources in the recent
2FHL catalogue (Ackermann et al. 2015) at energies higher than 50 GeV.
Thus, we extracted a sample of 190 clusters not related to previously known $\gamma$-ray
sources. 
One of them, however, was found to have a position very close to the famous GRB 130427A
\citep{maselli14} well detected also at these energies by \emph{Fermi}-LAT 
\citep{ackermann14}.

The matching of the positions of our 189 non-associated clusters with the subsample of
440 1WHSP sources gave 16 clusters with a positional coincidence within a maximum angular 
separation (computed using the cluster centroids' coordinates) of 6\arcmin\ (0\fdg1), 
comparable to the PSF radius at these energies (see also Paper I for a discussion on the matching radius).
The list of these sources and their most important cluster parameters are given in Table \ref{t:sourcelist}.  

It is interesting to note that if we double the maximum angular distance for the source 
matching to 12\arcmin, we find the same 16 sources, and this distance must be increased to 
16\arcmin\ to find one more matching source between the two samples. With the very large 
separation of 60\arcmin = 1\degr, 28 sources are found.
The number of sources with angular distance in the range [6\arcmin, 60\arcmin] is 12, less 
than those found within the original value of 6\arcmin.  
This simple test is a first significant indication to exclude the possibility that the 16 source matchings
are due to a simple random correspondence between the two samples and suggest a real physical 
association.
Some additional tests are presented in the following Subsections.

\subsection{MST clusters at energies higher than 7 GeV}

We performed a similar cluster search in the \emph{Fermi}-LAT sky at energies higher than 7 
GeV to confirm the $>$10 GeV clusters.
The analysis with a lower energy threshold is important because the higher percentage of 
background photons decreases the mean angular distances between them. 
Consequently, the clustering degree can decrease, unless the photon number in a 
cluster increases by an amount large enough to balance this effect.
We confirmed the detection of all the clusters, 14 of which with $M > 20$ (9 with 
$M > 30$).
The two clusters with $M < 20$ are MST~0241$-$3041 and MST~1033$+$3708
that are close to soft bright sources with a high number of photons.
However, when the MST is applied to regions excluding these sources their 
$M$ values result higher than 20 thus confirming the detection.  
Note also that above 7 GeV for the cluster MST~2040$-$4620 the angular 
distance between its centroid and the 1WHSP source position decreases to 
2\farcm35 above 7 GeV with the addition of only one photon.

\subsection{Statistical tests for the spatial association of the samples}

In order to verify the 16 associations of between the 1WSHP subsample and our 
MST clusters we developed some tests based on the search of spatial correspondences 
between the considered blazars with simulated samples having an equal number 
of elements (see also Paper I).

The expected number of correspondences $N_\mathrm{ex}$ due to random matching can be estimated 
from the ratio between the sum of the solid angles subtended by the clusters $\omega_i$ to the total 
solid angle of the explored sky region $\Omega$, multiplied by the number of blazars 
$N_b$:
\begin{equation}
N_\mathrm{ex} = N_b \frac{\sum_i \omega_i}{\Omega} = N_c N_b \frac{\omega}{\Omega}
\end{equation}
where $N_c$ is the number of clusters. We assume that all the clusters have the same solid angle $\omega$
corresponding to a circular region of radius equal to 0\fdg1.
From $\Omega = 7.256$ sr and $\omega = 9.57 \times 10 ^{-6}$ sr, with the number of
clusters in our sample ($N_c = 189$) and of blazars ($N_b = 440$) we obtain $N_\mathrm{ex} = 0.11$, 
a value two orders of magnitude smaller than our result.

Starting from our sample of 189 clusters we generated other samples of fake 
clusters sorting a position within an annular region centered at the true cluster 
centroid and having inner and outer radii of 0\fdg5 and 3\degr.
In this way we mantained the same large scale spatial distribution of clusters that 
could be affected by the inhomogeneities of the $\gamma$-ray background, due for 
instance to the distance from the Galactic belt.
From these simulations we found  
an average value of 0.085 matching clusters per run, rather close to the the previous 
estimate.
The small difference could be partially explained by the fraction of extracted fake
clusters can be located at Galactic latitudes lower than $|25\degr|$, and therefore
outside the region of interest.
As a consequence, the resulting Poissonian probability for finding 16 or more matching sources 
is of the order of $\sim$10$^{-31}$.
Similar probabilities are also obtained by shifting all the centroids' coordinates by 
the same quantities of the order of a fraction of degree.  
In conclusion, it is reasonable to assume that no more than one of the 16 associations 
with 1WHSP sources might be due to a random occurrence.

\begin{table*}[htb]
\caption{Standard unbinned likelihood analysis of the \emph{Fermi}-LAT data, see text for details. 
The third and fourth columns report photon fluxes in units of 10$^{-11}$ ph\,cm$^{-2}$\,s$^{-1}$.  
For sources below the usual significance threshold ($TS=25$) only upper limits are given.}
\centering
\label{t:tableTS}
{%\small
\begin{tabular}{lrccc}
\tableline
MST cluster &  $\sqrt{TS}$  &    Flux      &    Flux    & $\Gamma$  \\
            &               &  3--300 GeV  & 10--300 GeV  &  \\
\tableline                                                        
MST~0009$-$4316 &  5.66 & $5.2 \pm 1.9 $ 	& $1.5 \pm 0.8 $ & $2.0 \pm 0.4 $	\\
MST~0029$+$2053 &  6.12 & $4.5 \pm 1.7 $ 	& $2.0 \pm 0.9 $ & $1.6 \pm 0.3 $	\\
MST~0102$-$2002 &  5.57 & $4.3 \pm 1.6 $ 	& $1.9 \pm 0.9 $ & $1.6 \pm 0.3 $	\\
MST~0241$-$3041 &  3.62 & $\le 4.0 $ 	& $\le 0.8 $ 	   & ---	\\  %2.91
MST~0328$-$5716 &  4.88 & $\le 6.6 $ 	& $\le 1.2 $ 	   & ---	\\  %3.75
MST~0507$-$3346 &  8.34 & $8.6 \pm 2.2 $	& $3.3 \pm 1.1 $ & $1.8 \pm 0.2 $	\\
MST~0812$+$2821 &  5.19 & $4.7 \pm 1.7 $	& $1.9 \pm 0.9 $ & $1.7 \pm 0.3 $	\\
MST~1033$+$3708 &  6.41 & $6.3 \pm 2.0 $	& $2.3 \pm 0.9 $ & $1.8 \pm 0.3 $	\\
MST~1043$+$0054 &  7.50 & $8.0 \pm 2.3 $	& $3.7 \pm 1.2 $ & $1.6 \pm 0.2 $	\\
MST~1432$+$7644 &  4.15 & $\le 5.6     $	& $\le 1.0 $     & ---	\\  %2.60
MST~1550$-$0822 &  5.86 & $4.4 \pm 1.7 $	& $2.1 \pm 0.9 $ & $1.6 \pm 0.3 $	\\
MST~1623$+$0857 &  5.00 & $4.3 \pm 1.7 $   	& $1.5 \pm 0.7 $ & $1.9 \pm 0.4 $	\\
MST~1626$+$6300 &  7.79 & $8.1 \pm 2.2 $	& $3.2 \pm 1.1 $ & $1.7 \pm 0.3 $	\\
MST~1640$+$6852 &  6.07 & $6.7 \pm 2.3 $	& $2.1 \pm 0.9 $ & $1.9 \pm 0.4 $	\\
MST~2002$-$5736 &  7.25 & $12.1\pm 3.3 $	& $3.0 \pm 1.3 $ & $2.2 \pm 0.3 $	\\
MST~2040$-$4620 &  6.66 & $11.1\pm 2.9 $	& $2.3 \pm 1.1 $ & $2.3 \pm 0.4 $	\\
\hline
\end{tabular}
}
\end{table*}

\begin{figure}[h!]
\centering
\includegraphics[width=0.48\textwidth]{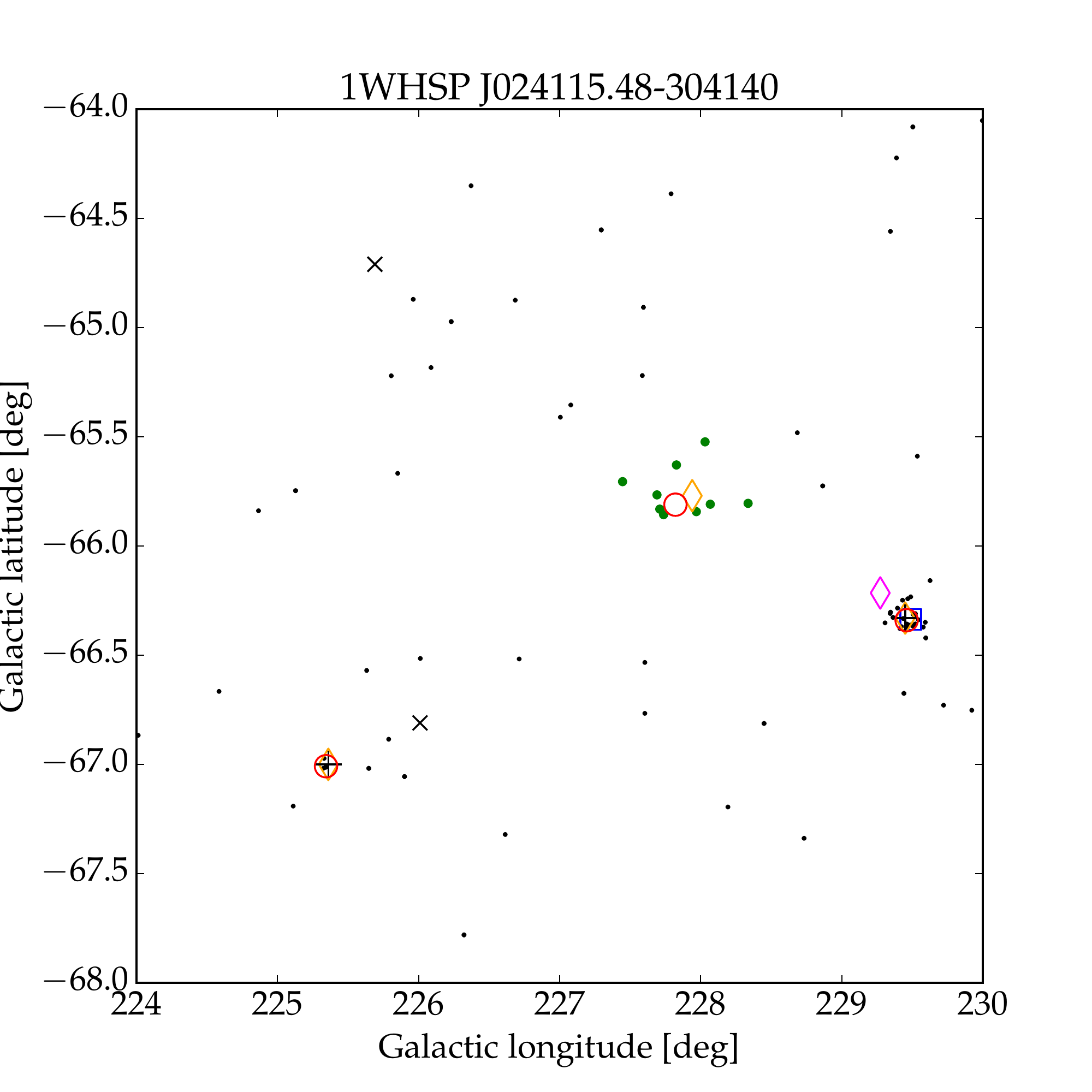}
\caption{Photon sky map in Galactic coordinates of the region around the BL Lac object
1WHSP J024115.48$-$304140. 
Black filled circles mark the photon coordinates at energies higher than 10 GeV, green points
are the photons in the cluster associated with the blazar;
large black crosses correspond to the optical coordinates of BL Lac objects in the 5Roma-BZCAT 
while small black X-symblos are Flat Spectrum Radio Quasar in the same catalogue, 
open orange diamonds are 1WHSP sources, blue open squares are the 3FGL sources, open red 
circles are the MST cluster positions, while the magenta diamonds are the sources in 
the D3PO catalogue \citep{selig14}.
}
\label{1WHSPmap}
\end{figure}

\section{Maximum likelihood analysis}

Adopting the same approach of Paper I, we performed also a standard unbinned likelihood 
analysis for each MST cluster.
A Region of Interest (ROI) of 10\degr\ radius was selected centered at the MST cluster 
centroid, and standard screening criteria were applied to the \emph{Fermi}-LAT data 
above 3 GeV. 
The likelihood analysis was performed considering all the 3FGL sources within 20\degr\ 
from the cluster centroid, as well as the Galactic and extragalactic diffuse emission. 
A further source with a power-law spectral distribution $F(E) = K E^{-\Gamma}$ was assumed at the MST coordinates. 
The normalization and spectral index of all the 3FGL sources within the ROI was allowed 
to vary in the fitting, while the parameters of the sources between 10\degr\ and 20\degr\ 
from the center of the field of view were fixed to their catalogue values.
From this analysis, we derived the likelihood Test Statistics ($TS$) and fluxes in the two 
3--300 and 10--300 GeV bands. 
The results are reported in Table~\ref{t:tableTS}: photon fluxes above 3 and 10 GeV between 10$^{-11}$ and 
10$^{-10}$ ph\,cm$^{-2}$\,s$^{-1}$ are obtained for the safe sources; 
for sources having a $TS$ value below the conventional significance threshold ($TS = 25$, 
corresponding to 5$\sigma$) upper limits are given. 

For the large majority of clusters the resulting $\sqrt{TS}$ are above the conventional
limit of 5.0 and their detection as genuine $\gamma$-ray sources is safely confirmed.
Only three clusters have lower values: two with $\sqrt{TS} > 4.0$ and only MST~0241$-$3041 
(associated with 1WHSP J024115.48$-$304140) has a value equal to 3.6.
Note that clusters with a low $\sqrt{TS}$ have also frequently a low $g$,
even if this correlation is not always verified because it depends on the different
energy ranges and on the local background.
There are, however, other indications to consider safe these three associations.

Figure~\ref{1WHSPmap} shows the LAT photon sky map at energies higher than 10 GeV in the
region around 1WHSP J024115.48$-$304140: there are four known 5BZCAT blazars, two of which are
Flat Spectrum Radio Quasars without any evidence of photon clusters in their 
neighbourhoods, while one of the two known BL Lac objects (5BZBJ0238$-$3116, 
3FGL J0238.4$-$3117) corresponds to the richest cluster in the field.
Both these two sources are in the 1WHSP catalogue, that also contains the source 
corresponding to the MST cluster of interest, MST~0241$-$3041.
This cluster has a rather low $g$ and some of the photons belonging to it are sparse.
A more refined analysis in a few degree wide field, excluding the bright source and
with a stronger cut length at $\Lambda_\mathrm{cut} = 0.5 \Lambda_{m}$, gives only the two
clusters corresponding to the 1WHSP sources: the one of interest with 7 photons,
$g = 3.27$, and $R_m = 0\farcm76$.
The other cluster, associated with a BL Lac object in the 5BZCAT and not previously
reported as $\gamma$-ray source, was already discussed in Paper I.

One of the two other clusters with $\sqrt{TS}$ lower than 5, MST 0328$-$5716, was firmly confirmed by the analysis at 
energies higher than 7 GeV that gave a cluster with 7 photons, $g = 3.88$ and $M = 27.17$.
In a further analysis above 3 GeV it resulted the cluster with the highest $g = 4.29$ 
in the region.
Moreover, at this lower energy threshold the angular separation between
the cluster centroid and the position of the 1WHSP source decreases to 1$'$.65.

The significance of the cluster MST 1432$+$7644 appears a bit weaker than the previous two.
In fact the analysis above 7 GeV with $\Lambda_\mathrm{cut} = 0.7 \Lambda_{m}$ gave a
rich cluster with 14 photons, but a rather low $g = 2.38$ and $M = 33.32$, while adopting
$\Lambda_\mathrm{cut} = 0.6 \Lambda_{m}$ the photon number was reduced to 6 with
$g = 3.08$ and $M = 18.48$.
This differences can be explained by the particular structure of this cluster with a high
density core embedded in a sparse halo so that a small change of $\Lambda_\mathrm{cut}$
would imply a large change in the cluster parameters.
The centroid position of the core is at about 5$'$.9 to the 1WHSP source coordinates while
that of the 14 photon cluster is at  1$'$.6.
It cannot be excluded that these differencies could contribute in giving a low $\sqrt{TS}$.
A more robust evidence for the existence of this emission could be obtained in the future by 
the richer statistics from a longer exposure.

We also searched if time distributions of events arriving from the regions of these 
sources were or not uniform suggesting the occurrence of possible brightening periods.  
Aperture photometry light curves of photons at energies above 3 GeV from a circular 
region of radius 0\fdg5 centered at the source position were extracted for all the clusters.
No significant variation or flaring activity was found for any source.

\section{Summary and discussion}

Our application of MST for searching localized photon clusters in the \emph{Fermi}-LAT sky
(Pass 8) confirmed the good performance of this method, particularly for detecting
clusters with a low number of photons.
We analysed sky images at energies higher than 10 GeV having rather low 
photon densities, allowing a robust detection of photon clusters having typical 
sizes comparable with the LAT PSF. 
In the selection procedure we adopted rather severe threshold values to reduce 
the possibility of spurious detections.

We found 189 clusters unassociated with 5BZCAT objects or previously reported 
$\gamma$-ray sources and the cross-matching with a subsample of 440 sources in 
the 1WHSP catalogues within a maximum angular separation of 6$'$ gave a list of 
16 clusters, further confirmed by a search above 7 GeV.
The significance of all the sources was then tested with the standard maximum likelihood 
analysis at energies higher than 3 GeV and the results confirmed the large majority
of these MST cluster detections.
Only 3 clusters resulted with a TS lower than 25 and only one of them below 16. 
Nevertheless, a further analysis on a small region provided additional arguments 
for their existence as emissions associated to 1WHSP sources.

It is interesting to note that the percentage of detected blazars in the entire 1WHSP 
subsample is quite low, about 5\%.
This can be explained either by a variable $\gamma$-ray emission with a low 
duty-cycle or by a low fraction of $\gamma$-ray loud sources, thus implying that
a selection based on low frequency data (radio, IR, optical and X) could be not 
very efficient to establish if sources will be bright also at the GeV to TeV energies. 
Note, however, that only two sources were found with a photon spectral index higher than
2.0, despite their relatively large uncertainties, indicating that the SED peak
is at particularly high energies as expected for HSP objects.

Five of these 16 objects were already reported as blazar candidates in previous surveys:
1WHSP J081231.25\-$+$282056 is in the sample of \cite{plotkin10}, based on SDSS
spectral data, while 1WHSP J104303.84\-$+$005420 is included in the ROXA list \citep{turriziani07}, and
1WHSP J000949.74\-$-$431650 is in the WIBRaLS \citep[][]{dabrusco14}
sample based on WISE colours, and the two sources 1WHSP~J0102\-50.93$-$200158 and 1WHSPJ162\-646.04$+$630048 
were inserted in the LORCAT catalogue of low-frequency flat spectrum 
radio sources by \cite{massaro14b}.

1WHSP sources classified as \emph{New} blazars are about the 28\% in the considered
subsample, but these are 44\% among our detected clusters.
This difference is not large enough to be significant (4 N-type sources expected against
7 found) but it could be an indication that a fraction of sources classified as
\emph{Candidates} might not be actual high energy emitters. 

\cite{arsioli15} introduced a figure of merit ($FOM$) that is higher for those 
1WHSP sources having a larger probability to be observed at TeV energies.
Sources associated with our clusters have all quite low $FOM$: the one with the 
highest value is 1WHSP J010250.93$-$200158 ($FOM = 0.40$).
In our 440 sources 1WHSP subsample the one with the highest $FOM$ is 1WHSP J092603.5+124333, 
for which it is equal to 1.58.
This is a galaxy at redshift $z = 0.187$ with an SDSS spectrum typical of an 
elliptical one without evidence of a particular nuclear activity and no high 
UV emission ($u - r = 3.7$ mag).
It has a radio flux density in FIRST of 27.5 mJy and is not detected in the X-ray
images by Swift-XRT. 
We searched for a possible $\gamma$-ray emission from this object but there is
no indication for any photon cluster in its surroundings.
It appears therefore a low luminosity radio galaxy instead of a blazar.

We searched also for a possible indication that the detected clusters are related to some
particular properties of the source, like the radio flux density, the WISE magnitude or
the synchrotron peak frequency in the SED (that is estimated in the 1WHSP). 
No significant correlation was found, indicating that these 16 objects have some specific property
to explain an enhanced high energy emission.

Our results are useful for investigating the nature and the population distribution
of low-brightness extragalactic high energy sources.
In particular, we have to distinguish between a rather nearby population of low luminosity
sources and that of high redshift objects, but the featureless continua of BL Lacs make this
goal quite hard.
It is therefore important to test if future data, based increasingly longer exposure
observations would provide a more robust evidence for describing the statistical properties
of these sources.

%______________________________________________________________
\begin{acknowledgements}
We acknowledge use of archival Fermi data. We made large use of the online version of the Roma-BZCAT 
and of the scientific tools developed at the ASI Science Data Center (ASDC),
 of the Sloan Digital Sky Survey (SDSS) archive, of the NED database and other astronomical 
catalogues distributed in digital form (Vizier and Simbad) at Centre de Dates astronomiques de 
Strasbourg (CDS) at the Louis Pasteur University.
\end{acknowledgements}
%______________________________________________________________
%\begin{thebibliography}

\bibliographystyle{spr-mp-nameyear-cnd}
\bibliography{bibliography} % your references Yourfile.bib 

\begin{thebibliography}{30}
% BibTex style file: spr-mp.bst (nameyear,cnd), 2011-05-27
\ifx \bisbn   \undefined \def \bisbn  #1{ISBN #1}\fi
\ifx \binits  \undefined \def \binits#1{#1} \fi
\ifx \bauthor  \undefined \def \bauthor#1{#1} \fi
\ifx \batitle  \undefined \def \batitle#1{#1} \fi
\ifx \bjtitle  \undefined \def \bjtitle#1{#1}\fi
\ifx \bvolume  \undefined \def \bvolume#1{\textbf{#1}}\fi
\ifx \byear  \undefined \def \byear#1{#1} \fi
\ifx \bissue  \undefined \def \bissue#1{#1} \fi
\ifx \bfpage  \undefined \def \bfpage#1{#1} \fi
\ifx \blpage  \undefined \def \blpage #1{#1} \fi
\ifx \burl  \undefined \def \burl#1{\textsf{#1}} \fi
\ifx \doiurl  \undefined \def \doiurl#1{\textsf{#1}} \fi
\ifx \betal  \undefined \def \betal{\textit{et al.}} \fi
\ifx \binstitute  \undefined \def \binstitute#1{#1} \fi
\ifx \binstitutionaled  \undefined \def \binstitutionaled#1{#1} \fi
\ifx \bctitle  \undefined \def \bctitle#1{#1} \fi
\ifx \beditor  \undefined \def \beditor#1{#1} \fi
\ifx \bpublisher  \undefined \def \bpublisher#1{#1} \fi
\ifx \bbtitle  \undefined \def \bbtitle#1{#1} \fi
\ifx \bedition  \undefined \def \bedition#1{#1} \fi
\ifx \bseriesno  \undefined \def \bseriesno#1{#1} \fi
\ifx \blocation  \undefined \def \blocation#1{#1} \fi
\ifx \bsertitle  \undefined \def \bsertitle#1{#1} \fi
\ifx \bsnm \undefined \def \bsnm#1{#1} \fi
\ifx \bsuffix \undefined \def \bsuffix#1{#1} \fi
\ifx \bparticle \undefined \def \bparticle#1{#1} \fi
\ifx \barticle \undefined \def \barticle#1{#1} \fi
\ifx \bconfdate \undefined \def \bconfdate #1{#1} \fi
\ifx \botherref \undefined \def \botherref #1{#1} \fi
\ifx \url \undefined \def \url#1{\textsf{#1}} \fi
\ifx \bchapter \undefined \def \bchapter#1{#1} \fi
\ifx \bbook \undefined \def \bbook#1{#1} \fi
\ifx \bcomment \undefined \def \bcomment#1{#1} \fi
\ifx \oauthor \undefined \def \oauthor#1{#1} \fi
\ifx \citeauthoryear \undefined \def \citeauthoryear#1{#1} \fi
\ifx \endbibitem  \undefined \def \endbibitem {}\fi
\ifx \bconflocation  \undefined \def \bconflocation#1{#1} \fi
\ifx \arxivurl  \undefined \def \arxivurl#1{\textsf{#1}} \fi

\bibitem[\protect\citeauthoryear{{Abdo} et~al.}{2010}]{abdo10}
\begin{barticle}
\bauthor{\bsnm{{Abdo}}, \binits{A.A.}},
\bauthor{\bsnm{{Ackermann}}, \binits{M.}},
\bauthor{\bsnm{{Ajello}}, \binits{M.}},
\bauthor{\bsnm{{Allafort}}, \binits{A.}},
\bauthor{\bsnm{{Antolini}}, \binits{E.}},
\bauthor{\bsnm{{Atwood}}, \binits{W.B.}},
\bauthor{\bsnm{{Axelsson}}, \binits{M.}},
\bauthor{\bsnm{{Baldini}}, \binits{L.}},
\bauthor{\bsnm{{Ballet}}, \binits{J.}},
\bauthor{\bsnm{{Barbiellini}}, \binits{G.}}, \betal:
\bjtitle{\apjs}
\bvolume{188},
\bfpage{405}
(\byear{2010})
\end{barticle}
\endbibitem

\bibitem[\protect\citeauthoryear{{Acero} et~al.}{2015}]{acero15}
\begin{barticle}
\bauthor{\bsnm{{Acero}}, \binits{F.}},
\bauthor{\bsnm{{Ackermann}}, \binits{M.}},
\bauthor{\bsnm{{Ajello}}, \binits{M.}},
\bauthor{\bsnm{{Albert}}, \binits{A.}},
\bauthor{\bsnm{{Atwood}}, \binits{W.B.}},
\bauthor{\bsnm{{Axelsson}}, \binits{M.}},
\bauthor{\bsnm{{Baldini}}, \binits{L.}},
\bauthor{\bsnm{{Ballet}}, \binits{J.}}, \betal:
\bjtitle{\apjs}
\bvolume{218},
\bfpage{23}
(\byear{2015})
\end{barticle}
\endbibitem

\bibitem[\protect\citeauthoryear{{Ackermann} et~al.}{2011}]{ackermann11}
\begin{barticle}
\bauthor{\bsnm{{Ackermann}}, \binits{M.}},
\bauthor{\bsnm{{Ajello}}, \binits{M.}},
\bauthor{\bsnm{{Allafort}}, \binits{A.}},
\bauthor{\bsnm{{Antolini}}, \binits{E.}},
\bauthor{\bsnm{{Atwood}}, \binits{W.B.}},
\bauthor{\bsnm{{Axelsson}}, \binits{M.}},
\bauthor{\bsnm{{Baldini}}, \binits{L.}},
\bauthor{\bsnm{{Ballet}}, \binits{J.}}, \betal:
\bjtitle{\apj}
\bvolume{743},
\bfpage{171}
(\byear{2011})
\end{barticle}
\endbibitem

\bibitem[\protect\citeauthoryear{{Ackermann} et~al.}{2013}]{ackermann13b}
\begin{barticle}
\bauthor{\bsnm{{Ackermann}}, \binits{M.}},
\bauthor{\bsnm{{Ajello}}, \binits{M.}},
\bauthor{\bsnm{{Allafort}}, \binits{A.}},
\bauthor{\bsnm{{Asano}}, \binits{K.}},
\bauthor{\bsnm{{Atwood}}, \binits{W.B.}},
\bauthor{\bsnm{{Baldini}}, \binits{L.}},
\bauthor{\bsnm{{Ballet}}, \binits{J.}},
\bauthor{\bsnm{{Barbiellini}}, \binits{G.}}, \betal:
\bjtitle{\apj}
\bvolume{765},
\bfpage{54}
(\byear{2013})
\end{barticle}
\endbibitem

\bibitem[\protect\citeauthoryear{{Ackermann} et~al.}{2014}]{ackermann14}
\begin{barticle}
\bauthor{\bsnm{{Ackermann}}, \binits{M.}},
\bauthor{\bsnm{{Ajello}}, \binits{M.}},
\bauthor{\bsnm{{Asano}}, \binits{K.}},
\bauthor{\bsnm{{Atwood}}, \binits{W.B.}},
\bauthor{\bsnm{{Axelsson}}, \binits{M.}},
\bauthor{\bsnm{{Baldini}}, \binits{L.}},
\bauthor{\bsnm{{Ballet}}, \binits{J.}},
\bauthor{\bsnm{{Barbiellini}}, \binits{G.}}, \betal:
\bjtitle{Science}
\bvolume{343},
\bfpage{42}
(\byear{2014})
\end{barticle}
\endbibitem

\bibitem[\protect\citeauthoryear{{Ackermann} et~al.}{2015}]{ackermann15}
\begin{barticle}
\bauthor{\bsnm{{Ackermann}}, \binits{M.}},
\bauthor{\bsnm{{Ajello}}, \binits{M.}},
\bauthor{\bsnm{{Atwood}}, \binits{W.B.}},
\bauthor{\bsnm{{Baldini}}, \binits{L.}},
\bauthor{\bsnm{{Ballet}}, \binits{J.}},
\bauthor{\bsnm{{Barbiellini}}, \binits{G.}},
\bauthor{\bsnm{{Bastieri}}, \binits{D.}}, \betal:
\bjtitle{\apj}
\bvolume{810},
\bfpage{14}
(\byear{2015})
\end{barticle}
\endbibitem

\bibitem[\protect\citeauthoryear{{Ackermann} et~al.}{2016}]{ackermann16}
\begin{barticle}
\bauthor{\bsnm{{Ackermann}}, \binits{M.}},
\bauthor{\bsnm{{Ajello}}, \binits{M.}},
\bauthor{\bsnm{{An}}, \binits{H.}},
\bauthor{\bsnm{{Baldini}}, \binits{L.}},
\bauthor{\bsnm{{Barbiellini}}, \binits{G.}},
\bauthor{\bsnm{{Bastieri}}, \binits{D.}},
\bauthor{\bsnm{{Bellazzini}}, \binits{R.}},
\bauthor{\bsnm{{Bissaldi}}, \binits{E.}}, \betal:
\bjtitle{\apj}
\bvolume{820},
\bfpage{72}
(\byear{2016})
\end{barticle}
\endbibitem

\bibitem[\protect\citeauthoryear{{Ajello} et~al.}{2015}]{ajello15}
\begin{botherref}
\oauthor{\bsnm{{Ajello}}, \binits{M.}},
\oauthor{\bsnm{{Gasparrini}}, \binits{D.}},
\oauthor{\bsnm{{S{\'a}nchez-Conde}}, \binits{M.}},
\oauthor{\bsnm{{Zaharijas}}, \binits{G.}},
\oauthor{\bsnm{{Gustafsson}}, \binits{M.}},
\oauthor{\bsnm{{Cohen-Tanugi}}, \binits{J.}},
\oauthor{\bsnm{{Dermer}}, \binits{C.D.}},
\oauthor{\bsnm{{Inoue}}, \binits{Y.}},
\oauthor{\bsnm{{Hartmann}}, \binits{D.}},
\oauthor{\bsnm{{Ackermann}}, \binits{M.}},
\oauthor{\bsnm{{Bechtol}}, \binits{K.}},
\oauthor{\bsnm{{Franckowiak}}, \binits{A.}},
\oauthor{\bsnm{{Reimer}}, \binits{A.}},
\oauthor{\bsnm{{Romani}}, \binits{R.W.}},
\oauthor{\bsnm{{Strong}}, \binits{A.W.}}:
{The Origin of the Extragalactic Gamma-Ray Background and Implications for Dark
  Matter Annihilation}
\textbf{800},
27
(2015)
\end{botherref}
\endbibitem

\bibitem[\protect\citeauthoryear{{Arsioli} et~al.}{2015}]{arsioli15}
\begin{barticle}
\bauthor{\bsnm{{Arsioli}}, \binits{B.}},
\bauthor{\bsnm{{Fraga}}, \binits{B.}},
\bauthor{\bsnm{{Giommi}}, \binits{P.}},
\bauthor{\bsnm{{Padovani}}, \binits{P.}},
\bauthor{\bsnm{{Marrese}}, \binits{P.M.}}:
\bjtitle{\aap}
\bvolume{579},
\bfpage{34}
(\byear{2015})
\end{barticle}
\endbibitem

\bibitem[\protect\citeauthoryear{{Atwood} et~al.}{2013}]{atwood13}
\begin{botherref}
\oauthor{\bsnm{{Atwood}}, \binits{W.}},
\oauthor{\bsnm{{Albert}}, \binits{A.}},
\oauthor{\bsnm{{Baldini}}, \binits{L.}},
\oauthor{\bsnm{{Tinivella}}, \binits{M.}},
\oauthor{\bsnm{{Bregeon}}, \binits{J.}},
\oauthor{\bsnm{{Pesce-Rollins}}, \binits{M.}},
\oauthor{\bsnm{{Sgr{\`o}}}, \binits{C.}},
\oauthor{\bsnm{{Bruel}}, \binits{P.}},
\oauthor{\bsnm{{Charles}}, \binits{E.}}, et al.:
ArXiv e-prints
(2013).
\arxivurl{1303.3514}
\end{botherref}
\endbibitem

\bibitem[\protect\citeauthoryear{{Campana} et~al.}{2016}]{campana16}
\begin{barticle}
\bauthor{\bsnm{{Campana}}, \binits{R.}},
\bauthor{\bsnm{{Massaro}}, \binits{E.}},
\bauthor{\bsnm{{Bernieri}}, \binits{E.}}:
\bjtitle{\apss}
\bvolume{in press},
(\byear{2016})
\end{barticle}
\endbibitem

\bibitem[\protect\citeauthoryear{{Campana} et~al.}{2008}]{campana08}
\begin{barticle}
\bauthor{\bsnm{{Campana}}, \binits{R.}},
\bauthor{\bsnm{{Massaro}}, \binits{E.}},
\bauthor{\bsnm{{Gasparrini}}, \binits{D.}},
\bauthor{\bsnm{{Cutini}}, \binits{S.}},
\bauthor{\bsnm{{Tramacere}}, \binits{A.}}:
\bjtitle{\mnras}
\bvolume{383},
\bfpage{1166}
(\byear{2008})
\end{barticle}
\endbibitem

\bibitem[\protect\citeauthoryear{{Campana} et~al.}{2013}]{campana13}
\begin{barticle}
\bauthor{\bsnm{{Campana}}, \binits{R.}},
\bauthor{\bsnm{{Bernieri}}, \binits{E.}},
\bauthor{\bsnm{{Massaro}}, \binits{E.}},
\bauthor{\bsnm{{Tinebra}}, \binits{F.}},
\bauthor{\bsnm{{Tosti}}, \binits{G.}}:
\bjtitle{\apss}
\bvolume{347},
\bfpage{169}
(\byear{2013})
\end{barticle}
\endbibitem

\bibitem[\protect\citeauthoryear{{Campana} et~al.}{2015}]{campana15}
\begin{barticle}
\bauthor{\bsnm{{Campana}}, \binits{R.}},
\bauthor{\bsnm{{Massaro}}, \binits{E.}},
\bauthor{\bsnm{{Bernieri}}, \binits{E.}},
\bauthor{\bsnm{{D'Amato}}, \binits{Q.}}:
\bjtitle{\apss}
\bvolume{360},
\bfpage{65}
(\byear{2015})
\end{barticle}
\endbibitem

\bibitem[\protect\citeauthoryear{{Cutri} et~al.}{2013}]{cutri13}
\begin{botherref}
\oauthor{\bsnm{{Cutri}}, \binits{R.M.}},
\oauthor{\bsnm{{Wright}}, \binits{E.L.}},
\oauthor{\bsnm{{Conrow}}, \binits{T.}},
\oauthor{\bsnm{{Fowler}}, \binits{J.W.}},
\oauthor{\bsnm{{Eisenhardt}}, \binits{P.R.M.}},
\oauthor{\bsnm{{Grillmair}}, \binits{C.}},
\oauthor{\bsnm{{Kirkpatrick}}, \binits{J.D.}},
\oauthor{\bsnm{{Masci}}, \binits{F.}}, et al.:
{Explanatory Supplement to the AllWISE Data Release Products}.
Technical report,
(November 2013)
\end{botherref}
\endbibitem

\bibitem[\protect\citeauthoryear{{D'Abrusco} et~al.}{2014}]{dabrusco14}
\begin{barticle}
\bauthor{\bsnm{{D'Abrusco}}, \binits{R.}},
\bauthor{\bsnm{{Massaro}}, \binits{F.}},
\bauthor{\bsnm{{Paggi}}, \binits{A.}},
\bauthor{\bsnm{{Smith}}, \binits{H.A.}},
\bauthor{\bsnm{{Masetti}}, \binits{N.}},
\bauthor{\bsnm{{Landoni}}, \binits{M.}},
\bauthor{\bsnm{{Tosti}}, \binits{G.}}:
\bjtitle{\apjs}
\bvolume{215},
\bfpage{14}
(\byear{2014})
\end{barticle}
\endbibitem

\bibitem[\protect\citeauthoryear{{Giommi} et~al.}{1999}]{giommi99}
\begin{barticle}
\bauthor{\bsnm{{Giommi}}, \binits{P.}},
\bauthor{\bsnm{{Menna}}, \binits{M.T.}},
\bauthor{\bsnm{{Padovani}}, \binits{P.}}:
\bjtitle{\mnras}
\bvolume{310},
\bfpage{465}
(\byear{1999})
\end{barticle}
\endbibitem

\bibitem[\protect\citeauthoryear{{Giommi} et~al.}{2005}]{giommi05}
\begin{barticle}
\bauthor{\bsnm{{Giommi}}, \binits{P.}},
\bauthor{\bsnm{{Piranomonte}}, \binits{S.}},
\bauthor{\bsnm{{Perri}}, \binits{M.}},
\bauthor{\bsnm{{Padovani}}, \binits{P.}}:
\bjtitle{\aap}
\bvolume{434},
\bfpage{385}
(\byear{2005})
\end{barticle}
\endbibitem

\bibitem[\protect\citeauthoryear{{Maselli} et~al.}{2014}]{maselli14}
\begin{barticle}
\bauthor{\bsnm{{Maselli}}, \binits{A.}},
\bauthor{\bsnm{{Melandri}}, \binits{A.}},
\bauthor{\bsnm{{Nava}}, \binits{L.}},
\bauthor{\bsnm{{Mundell}}, \binits{C.G.}},
\bauthor{\bsnm{{Kawai}}, \binits{N.}},
\bauthor{\bsnm{{Campana}}, \binits{S.}},
\bauthor{\bsnm{{Covino}}, \binits{S.}},
\bauthor{\bsnm{{Cummings}}, \binits{J.R.}}, \betal:
\bjtitle{Science}
\bvolume{343},
\bfpage{48}
(\byear{2014})
\end{barticle}
\endbibitem

\bibitem[\protect\citeauthoryear{{Massaro} et~al.}{2014}]{massaro14}
\begin{bbook}
\bauthor{\bsnm{{Massaro}}, \binits{E.}},
\bauthor{\bsnm{{Maselli}}, \binits{A.}},
\bauthor{\bsnm{{Leto}}, \binits{C.}}, \betal:
\bbtitle{Multifrequency Catalogue of Blazars},
\bedition{5th} edn.
\bpublisher{Aracne Editrice},
\blocation{Rome}
(\byear{2014})
\end{bbook}
\endbibitem

\bibitem[\protect\citeauthoryear{{Massaro} et~al.}{2015}]{massaro15}
\begin{barticle}
\bauthor{\bsnm{{Massaro}}, \binits{E.}},
\bauthor{\bsnm{{Maselli}}, \binits{A.}},
\bauthor{\bsnm{{Leto}}, \binits{C.}},
\bauthor{\bsnm{{Marchegiani}}, \binits{P.}},
\bauthor{\bsnm{{Perri}}, \binits{M.}},
\bauthor{\bsnm{{Giommi}}, \binits{P.}},
\bauthor{\bsnm{{Piranomonte}}, \binits{S.}}:
\bjtitle{\apss}
\bvolume{357},
\bfpage{75}
(\byear{2015})
\end{barticle}
\endbibitem

\bibitem[\protect\citeauthoryear{{Massaro} et~al.}{2016}]{massaro16}
\begin{barticle}
\bauthor{\bsnm{{Massaro}}, \binits{F.}},
\bauthor{\bsnm{{Thompson}}, \binits{D.J.}},
\bauthor{\bsnm{{Ferrara}}, \binits{E.C.}}:
\bjtitle{\aapr}
\bvolume{24},
\bfpage{2}
(\byear{2016})
\end{barticle}
\endbibitem

\bibitem[\protect\citeauthoryear{{Massaro} et~al.}{2011}]{massaro11}
\begin{barticle}
\bauthor{\bsnm{{Massaro}}, \binits{F.}},
\bauthor{\bsnm{{D'Abrusco}}, \binits{R.}},
\bauthor{\bsnm{{Ajello}}, \binits{M.}},
\bauthor{\bsnm{{Grindlay}}, \binits{J.E.}},
\bauthor{\bsnm{{Smith}}, \binits{H.A.}}:
\bjtitle{\apjl}
\bvolume{740},
\bfpage{48}
(\byear{2011})
\end{barticle}
\endbibitem

\bibitem[\protect\citeauthoryear{{Massaro} et~al.}{2013}]{massaro13}
\begin{barticle}
\bauthor{\bsnm{{Massaro}}, \binits{F.}},
\bauthor{\bsnm{{Paggi}}, \binits{A.}},
\bauthor{\bsnm{{Errando}}, \binits{M.}},
\bauthor{\bsnm{{D'Abrusco}}, \binits{R.}},
\bauthor{\bsnm{{Masetti}}, \binits{N.}},
\bauthor{\bsnm{{Tosti}}, \binits{G.}},
\bauthor{\bsnm{{Funk}}, \binits{S.}}:
\bjtitle{\apjs}
\bvolume{207},
\bfpage{16}
(\byear{2013})
\end{barticle}
\endbibitem

\bibitem[\protect\citeauthoryear{{Massaro} et~al.}{2014}]{massaro14b}
\begin{barticle}
\bauthor{\bsnm{{Massaro}}, \binits{F.}},
\bauthor{\bsnm{{Giroletti}}, \binits{M.}},
\bauthor{\bsnm{{D'Abrusco}}, \binits{R.}},
\bauthor{\bsnm{{Masetti}}, \binits{N.}},
\bauthor{\bsnm{{Paggi}}, \binits{A.}},
\bauthor{\bsnm{{Cowperthwaite}}, \binits{P.S.}},
\bauthor{\bsnm{{Tosti}}, \binits{G.}},
\bauthor{\bsnm{{Funk}}, \binits{S.}}:
\bjtitle{\apjs}
\bvolume{213},
\bfpage{3}
(\byear{2014})
\end{barticle}
\endbibitem

\bibitem[\protect\citeauthoryear{{Nolan} et~al.}{2012}]{nolan12}
\begin{barticle}
\bauthor{\bsnm{{Nolan}}, \binits{P.L.}},
\bauthor{\bsnm{{Abdo}}, \binits{A.A.}},
\bauthor{\bsnm{{Ackermann}}, \binits{M.}},
\bauthor{\bsnm{{Ajello}}, \binits{M.}},
\bauthor{\bsnm{{Allafort}}, \binits{A.}},
\bauthor{\bsnm{{Antolini}}, \binits{E.}},
\bauthor{\bsnm{{Atwood}}, \binits{W.B.}},
\bauthor{\bsnm{{Axelsson}}, \binits{M.}},
\bauthor{\bsnm{{Baldini}}, \binits{L.}},
\bauthor{\bsnm{{Ballet}}, \binits{J.}},
\bauthor{\bparticle{et} \bsnm{al.}}:
\bjtitle{\apjs}
\bvolume{199},
\bfpage{31}
(\byear{2012})
\end{barticle}
\endbibitem

\bibitem[\protect\citeauthoryear{{Plotkin} et~al.}{2010}]{plotkin10}
\begin{barticle}
\bauthor{\bsnm{{Plotkin}}, \binits{R.M.}},
\bauthor{\bsnm{{Anderson}}, \binits{S.F.}},
\bauthor{\bsnm{{Brandt}}, \binits{W.N.}},
\bauthor{\bsnm{{Diamond-Stanic}}, \binits{A.M.}},
\bauthor{\bsnm{{Fan}}, \binits{X.}},
\bauthor{\bsnm{{Hall}}, \binits{P.B.}},
\bauthor{\bsnm{{Kimball}}, \binits{A.E.}},
\bauthor{\bsnm{{Richmond}}, \binits{M.W.}},
\bauthor{\bsnm{{Schneider}}, \binits{D.P.}},
\bauthor{\bsnm{{Shemmer}}, \binits{O.}},
\bauthor{\bsnm{{Voges}}, \binits{W.}},
\bauthor{\bsnm{{York}}, \binits{D.G.}},
\bauthor{\bsnm{{Bahcall}}, \binits{N.A.}},
\bauthor{\bsnm{{Snedden}}, \binits{S.}},
\bauthor{\bsnm{{Bizyaev}}, \binits{D.}},
\bauthor{\bsnm{{Brewington}}, \binits{H.}},
\bauthor{\bsnm{{Malanushenko}}, \binits{V.}},
\bauthor{\bsnm{{Malanushenko}}, \binits{E.}},
\bauthor{\bsnm{{Oravetz}}, \binits{D.}},
\bauthor{\bsnm{{Pan}}, \binits{K.}},
\bauthor{\bsnm{{Simmons}}, \binits{A.}}:
\bjtitle{\aj}
\bvolume{139},
\bfpage{390}
(\byear{2010})
\end{barticle}
\endbibitem

\bibitem[\protect\citeauthoryear{{Selig} et~al.}{2014}]{selig14}
\begin{botherref}
\oauthor{\bsnm{{Selig}}, \binits{M.}},
\oauthor{\bsnm{{Vacca}}, \binits{V.}},
\oauthor{\bsnm{{Oppermann}}, \binits{N.}},
\oauthor{\bsnm{{En{\ss}lin}}, \binits{T.A.}}:
ArXiv e-prints
(2014).
\arxivurl{1410.4562}
\end{botherref}
\endbibitem

\bibitem[\protect\citeauthoryear{{Turriziani} et~al.}{2007}]{turriziani07}
\begin{barticle}
\bauthor{\bsnm{{Turriziani}}, \binits{S.}},
\bauthor{\bsnm{{Cavazzuti}}, \binits{E.}},
\bauthor{\bsnm{{Giommi}}, \binits{P.}}:
\bjtitle{\aap}
\bvolume{472},
\bfpage{699}
(\byear{2007})
\end{barticle}
\endbibitem

\bibitem[\protect\citeauthoryear{{Wright} et~al.}{2010}]{wright10}
\begin{barticle}
\bauthor{\bsnm{{Wright}}, \binits{E.L.}},
\bauthor{\bsnm{{Eisenhardt}}, \binits{P.R.M.}},
\bauthor{\bsnm{{Mainzer}}, \binits{A.K.}},
\bauthor{\bsnm{{Ressler}}, \binits{M.E.}},
\bauthor{\bsnm{{Cutri}}, \binits{R.M.}},
\bauthor{\bsnm{{Jarrett}}, \binits{T.}},
\bauthor{\bsnm{{Kirkpatrick}}, \binits{J.D.}},
\bauthor{\bsnm{{Padgett}}, \binits{D.}},
\bauthor{\bsnm{{McMillan}}, \binits{R.S.}},
\bauthor{\bsnm{{Skrutskie}}, \binits{M.}},
\bauthor{\bsnm{{Stanford}}, \binits{S.A.}},
\bauthor{\bsnm{{Cohen}}, \binits{M.}},
\bauthor{\bsnm{{Walker}}, \binits{R.G.}},
\bauthor{\bsnm{{Mather}}, \binits{J.C.}},
\bauthor{\bsnm{{Leisawitz}}, \binits{D.}},
\bauthor{\bsnm{{Gautier}}, \binits{T.N.} \bsuffix{III}},
\bauthor{\bsnm{{McLean}}, \binits{I.}},
\bauthor{\bsnm{{Benford}}, \binits{D.}},
\bauthor{\bsnm{{Lonsdale}}, \binits{C.J.}},
\bauthor{\bsnm{{Blain}}, \binits{A.}},
\bauthor{\bsnm{{Mendez}}, \binits{B.}},
\bauthor{\bsnm{{Irace}}, \binits{W.R.}},
\bauthor{\bsnm{{Duval}}, \binits{V.}},
\bauthor{\bsnm{{Liu}}, \binits{F.}},
\bauthor{\bsnm{{Royer}}, \binits{D.}},
\bauthor{\bsnm{{Heinrichsen}}, \binits{I.}},
\bauthor{\bsnm{{Howard}}, \binits{J.}},
\bauthor{\bsnm{{Shannon}}, \binits{M.}},
\bauthor{\bsnm{{Kendall}}, \binits{M.}},
\bauthor{\bsnm{{Walsh}}, \binits{A.L.}},
\bauthor{\bsnm{{Larsen}}, \binits{M.}},
\bauthor{\bsnm{{Cardon}}, \binits{J.G.}},
\bauthor{\bsnm{{Schick}}, \binits{S.}},
\bauthor{\bsnm{{Schwalm}}, \binits{M.}},
\bauthor{\bsnm{{Abid}}, \binits{M.}},
\bauthor{\bsnm{{Fabinsky}}, \binits{B.}},
\bauthor{\bsnm{{Naes}}, \binits{L.}},
\bauthor{\bsnm{{Tsai}}, \binits{C.-W.}}:
\bjtitle{\aj}
\bvolume{140},
\bfpage{1868}
(\byear{2010})
\end{barticle}
\endbibitem

\end{thebibliography}

%\end{thebibliography}
%______________________________________________________________
\end{document}